# The DIME Architecture: A Unified Operational Algorithm for Neural Representation, Dynamics, Control and Integration

Ionel Cristian Vladu[1]*, Nicu George Bîzdoacă[2], Ionica Pirici[3], Tudor-Adrian Bălşeanu[4], Eduard Nicuşor Bondoc[5]

[1] Department of Electromechanics, Environment and Applied Informatics, Faculty of Electrical Engineering, University of Craiova, 200440 Craiova, Romania

[2] Department of Mechatronics and Robotics, Faculty of Automation, Computers and Electronics, University of Craiova, 200440 Craiova, Romania

[3] Department of Human Anatomy, University of Medicine and Pharmacy of Craiova, 200349 Craiova, Romania

[4] Department of Physiology, University of Medicine and Pharmacy of Craiova, 200349 Craiova, Romania

[5] Clinical Psychiatrist, Neuropsychiatry Hospital of Craiova; Associate Researcher, Department of Neuroscience, University of Medicine and Pharmacy of Craiova, 200349 Craiova, Romania

*Corresponding Author: cristian.vladu@edu.ucv.ro

## Author's note.

This arXiv version is a condensed architectural synopsis of the DIME framework. A more extensive theoretical treatment—including formal definitions, architectural closure, biological grounding across scales, falsifiable predictions, algorithmic specifications, and simulation results—is provided in the companion monograph [1]. The present document introduces the core operational commitments of the framework, while the monograph develops their full theoretical, biological, and computational consequences.

## Abstract

Modern neuroscience has accumulated a large body of evidence on perception, memory, prediction, valuation and consciousness, yet still lacks an explicit operational architecture capable of integrating these phenomena within a single computational framework [2], [3], [4], [5], [6], [7]. Existing theories typically address isolated aspects of neural function: predictive coding and active inference emphasize hierarchical inference and prediction error minimization [8], [9], engram theories describe memory storage and retrieval through distributed cell assemblies and pattern completion mechanisms [10], [11], [12], neuromodulatory accounts focus on value-dependent modulation of plasticity and behaviour [13], [14], [15], while global workspace and large-scale network models investigate the integration underlying conscious access [7], [16], [17]. Despite their explanatory power, these approaches are typically developed within partially overlapping but not formally unified architectural schemes.

This work introduces DIME (Detect–Integrate–Mark–Execute), an integrative neural architecture in which perception, memory, valuation and conscious access are described within a single operational loop, without claiming exhaustiveness or exclusivity. The architecture is built on four interacting components: engrams, defined as distributed recurrent neural structures supporting multiple activation trajectories rather than static memory traces [10], [11], [12], [15]; execution threads, defined as spatiotemporal trajectories implementing neural processes across time and modalities [18], [19], [20]; marker systems, corresponding to neuromodulatory and limbic mechanisms that implement value-based selection and control, governing gain modulation, plasticity gating, trajectory competition, and stabilization [13], [14], [15], [21], [22]; and hyperengrams, large-scale, temporally stabilized and value-marked integrative states associated with operational conscious access [16], [7], [23], [17], [24], [25].

The DIME framework is biologically grounded and consistent with empirical evidence from hippocampal indexing, cortical–subcortical loops, recurrent processing, replay phenomena, large-scale network integration and neuromodulatory regulation [10], [7], [11], [18], [21], [26]. At the same time, it is formulated at an abstract computational level that makes it applicable to artificial intelligence and robotics, offering an architectural template for agents in which representation, valuation and temporal sequencing emerge from the same underlying mechanism, rather than from separate modules or externally imposed reward structures [27], [28], [29].

This manuscript is intended as an architectural synopsis of the DIME (Detect–Integrate–Mark–Execute) framework, designed to introduce its core operational commitments and explanatory scope. A more extensive theoretical exposition—including formalization, neurobiological grounding, simulations, and applications—is available in the companion monograph on Zenodo [1].

## 1. Introduction

Understanding how the brain constructs perception, memory, adaptive behaviour and conscious experience remains a central challenge in contemporary neuroscience. While experimental progress has been substantial across multiple scales — from synaptic plasticity [30], [31] to cortical microcircuits [32], large-scale networks [7], [23] and neuromodulatory systems [13] — these findings coexist without a unifying architecture connecting them into a single operational principle.

Most existing frameworks address isolated aspects of cognition. Predictive coding and active inference describe perception as hierarchical prediction error minimization [8], [9]. Engram theories account for memory using distributed cell assemblies capable of pattern completion and replay [10], [11]. Neuromodulatory accounts highlight how dopamine, noradrenaline, acetylcholine and serotonin modulate plasticity, salience and behavioural policies [13], [14]. Global workspace theories emphasize large-scale broadcasting mechanisms associated with conscious access [16], [17]. Although individually influential, these frameworks do not converge into a unified description of how perception, memory, valuation, planning and consciousness relate to a common computational cycle.

The DIME architecture proposes such a unification by distinguishing four complementary aspects of neural organization:

1. **engrams**, the structural substrate functioning as dynamic neural programs;
2. **execution threads**, the process-level trajectories that implement cognition in time;
3. **marker systems**, the control layer assigning value, relevance and goal-dependence;

4. **hyperengrams**, the level of global integration linking identity, context, episodic history and goals.

Within this framework, the brain can be interpreted not primarily as a collection of independent modules for perception, memory or executive control, but as a single recurrent system whose functional distinctions emerge from the same underlying operational cycle. Instead, cognition emerges from a continuous cycle of **Detect–Integrate–Mark–Execute**, in which engrame activation, thread propagation, marker modulation and hyperengram integration interact synergistically across scales.

A more detailed exposition of this architecture—including its formal closure, biological instantiation across scales, pathological implications, and computational realizations—is provided in the companion monograph [1]. The present preprint develops the conceptual and architectural foundations of the framework.

### 1.1 Pre-rational selection and neuromodulatory control

Neural development suggests that behavioural selection precedes rational deliberation. Early cognition is governed not by explicit cortical reasoning, but by neuromodulatory valuation systems regulating plasticity and trajectory stabilization. This supports the hypothesis that value-based selection of neural trajectories – mediated by dopamine, noradrenaline, acetylcholine and limbic circuits – constitutes the primary control mechanism out of which rational deliberation subsequently emerges.

### 1.2 Duality of FI/FM flows

This observation underlies the dual flow structure of DIME: a representational stream (execution thread propagation across engrams) and a modulatory stream (marker field shaping gain, exploration and plasticity). Whereas most cognitive models treat value signals as optional reinforcement, DIME treats modulatory dynamics as intrinsic computational parameters governing thread evolution.

### 1.3 Threads as programs; engrams as executable structures

Within this framework, execution threads function as neural programs, not metaphors: behaviour arises from structured trajectories across engram networks, whose continuation, branching or suppression is determined by marker fields. Thus, engrams provide representational substrates while threads constitute executable processes — a unification rarely articulated explicitly in current theory.

This paper does not aim to present a complete mathematical model or experimental validation; its purpose is to formalize the architectural and algorithmic core of DIME.

## 2. THE NEED FOR A UNIFIED NEURAL ARCHITECTURE

Contemporary neuroscience offers detailed descriptions at microscopic (synaptic), mesoscopic (circuit), and macroscopic (network) levels [2], [7], [30], [23], [31], [32]. However, these descriptions do not amount to a unified architecture capable of expressing how different cognitive phenomena arise from the same underlying mechanism. Traditional psychological distinctions — episodic memory, semantic knowledge, procedural skills, emotion, attention, executive control — map only loosely onto neural circuitry, where the same neurons and regions participate in multiple functions depending on context [33], [34].

The lack of a unifying framework leads to three main limitations. First, it obscures the identification of a fundamental unit of representation and computation. Most theories treat neurons or layers as basic units, despite evidence for distributed assemblies, pattern completion and multiscale integration [10], [11], [12].

Second, it results in redundant explanations: separate theories for memory consolidation, attentional modulation, decision-making or emotion regulation, even though these processes share overlapping neural mechanisms [15].

Third, without a common architecture, computational models cannot reproduce the interdependence of perception, memory, action and valuation observed in biological systems [27]. What may be needed is an intermediate level of description — abstract enough to unify diverse phenomena, yet constrained enough to respect neurobiology. Such a description must identify:

(1) a structural unit of representation beyond the individual neuron;

(2) a dynamic unit describing how neural activity propagates;

(3) a control mechanism assigning value and modulating trajectories;

(4) a global integrative substrate capable of supporting conscious experience.

The DIME architecture addresses these requirements by grounding cognition in the interaction among engrams, execution threads, marker systems and hyperengrams [1].

## 2.1 Lack of a Unified Structural Unit

The search for a fundamental unit of neural representation has historically focused on the individual neuron, conceptualized either as a feature detector or as a weighted integrator of synaptic inputs [30], [31], [32]. While this view has been instrumental in computational modeling, it does not reflect the distributed nature of neural encoding observed in biological systems. Experimental studies demonstrate that percepts, memories and decisions are supported not by single neurons, but by **assemblies of neurons** whose coordinated activity forms stable, reactivatable patterns known as cell assemblies or engrams [10], [11], [12].

These assemblies exhibit properties that single neurons cannot:

- **pattern completion**, where partial cues trigger full representational states [10];
- **pattern separation**, particularly in the hippocampus [35];
- **context-dependent reconfiguration**, where the same neurons participate in different cognitive functions depending on global state [33], [34];
- **multiscale recurrence**, combining local microcircuit recurrency with long-range cortico–hippocampal loops [23], [18].

Such findings argue for a **structural unit** above the neuron: a distributed, recurrent network capable of encoding multiple potential trajectories of activation. In the DIME framework, this structural unit is the **engram**, defined not only by its connectivity but by its dynamic repertoire of possible activations.

Without adopting such a unit, theories of perception, memory and decision-making remain fragmented, each implicitly appealing to different structural abstractions that fail to generalize across cognitive domains [4], [5].

## 2.2 Lack of a Unified Process Unit

Beyond structure, cognition requires a **processual unit** describing how neural activity flows in time. Existing theories implicitly rely on several mutually incompatible proposals: single-neuron spike trains, attractor dynamics, state-space trajectories, and predictive-error hierarchies [8], [9], [12], [36]. The absence of a common process unit makes it difficult to map biological dynamics onto computational models in a consistent manner.

Empirical evidence shows that neural activity unfolds as **coherent spatiotemporal trajectories**, linking assemblies across cortical and subcortical regions [11], [18], [19]. These trajectories appear in perception (feedforward–feedback sweeps), memory recall (hippocampal–cortical replay), planning (prefrontal routing of sensorimotor sequences), and even spontaneous thought (default mode transitions). Such trajectories are neither static attractors nor pure feedforward cascades; they are **dynamic, context-modulated paths** through the space of possible engram activations.

In DIME, the relevant process unit is the **execution thread**: a temporally extended, causally coherent sequence of activations traversing engrame networks. Execution threads unify mechanisms that appear disparate when described separately: perceptual flow, memory retrieval, simulation, motor planning and inner speech.

Without a unified process unit, neural theories cannot explain the continuity of cognition, the integration of multimodal content, or the temporal coherence of conscious experience [16], [7].

### 2.3 Lack of a Unified Mechanism of Value and Control

A central challenge for any neural architecture is explaining how content-independent mechanisms assign value, relevance and priority to representations. Existing accounts fragment this problem across: dopaminergic reinforcement learning [8], attentional theories of salience [17], affective modulation by limbic structures [21], and executive control theories in prefrontal cortex [15].

However, value assignment is a **global constraint** that shapes nearly every aspect of cognition:

- It selects which percepts dominate awareness;
- It determines which memories are consolidated or suppressed;
- It modulates learning rates and prediction thresholds;
- It governs the direction and persistence of behavioural policies.

Neuromodulatory systems — dopamine, serotonin, noradrenaline, acetylcholine — operate across vast spatial scales, modulating synaptic gain, excitability and plasticity [13], [14], [21]. Limbic structures such as the amygdala and hippocampus contribute affective and contextual weighting [10], [22].

Despite their diversity, these systems can be interpreted as forming a **unified control layer**, assigning value and relevance to representational states. In DIME, this layer is formalized as **marker systems**: distributed mechanisms that regulate which engrams and execution threads are amplified, inhibited, stabilized or reconfigured.

Without a unified control mechanism, theories struggle to explain how emotion, attention, motivation and learning interact to shape behaviour and subjective experience.

Although the core DIME operational cycle is present early in development, the space of valuation and control through which it operates expands progressively during ontogeny. Early postnatal cognition is dominated by primary markers linked to homeostasis, nociception, reward, and basic avoidance or approach behaviors. Over development, these primary markers are integrated into increasingly complex and abstract configurations, supporting social evaluation, goal persistence, self-regulation, and long-term planning. This maturation does not involve the introduction of new processing mechanisms, but an expansion in the dimensionality and integration of the marker space that modulates execution-thread selection. Within this framework, heightened emotional reactivity in early life and the delayed maturation of executive control emerge naturally from the asynchronous development of marker systems operating over an otherwise stable operational cycle.

### 2.4 Lack of a Scalable Framework for Conscious Integration

Consciousness remains one of the most challenging dimensions for unification. Existing approaches — global workspace theory [16], recurrent processing theory [26], integrated information theory [24], and large-scale network models [7], [23] — illuminate important aspects but lack convergence regarding the underlying substrate and operational mechanism.

Neural evidence indicates that conscious experience:

- requires **large-scale integration** across cortical and subcortical networks [16];
- depends on **sustained, recurrent activation** rather than feedforward sweeps [26];
- displays a **temporal continuity** shaped by memory, expectation and action plans [25];
- is strongly modulated by affective and motivational states [15].

What is missing is an account that links:

- representational content (engrams),
- temporal sequencing (execution threads),
- value modulation (marker systems),
- and global integration (the conscious field).

In the DIME framework, conscious access is operationally associated with the formation of a **hyperengram**: a large-scale, temporally stabilized and value-marked integrative state. This definition is functional rather than ontological. This interpretation is compatible with data on cortical broadcasting, hippocampal indexing, prefrontal coordination and network-level coherence [16], [7], [23], [22].

A framework lacking such a global integrative substrate may face difficulties scaling from simple sensorimotor control to the unified, temporally extended conscious experience characteristic of human cognition.

## 3. CORE COMPONENTS OF THE DIME ARCHITECTURE

The DIME architecture is grounded on four complementary components that jointly implement the fundamental operations of cognition: engrams, execution threads, marker systems and hyperengrams. Each component captures a different dimension of neural computation—structure, process, control and global integration—and none can be reduced to the others. Their interaction provides the substrate for the Detect–Integrate–Mark–Execute cycle described throughout the paper and fully elaborated in the companion monograph [1].

### 3.1 Engrams: Dynamic Neural Programs

The term *engram* traditionally refers to a physical substrate of memory, typically conceptualized as a stable pattern of synaptic connectivity encoding a specific experience [10]. Recent advances, however, demonstrate that engrams are **not static storage units** but dynamic networks capable of multiple activation patterns, reconsolidation and context-dependent reconfiguration [11], [12], [37].

In the DIME framework, an engram is defined as a **distributed recurrent network encoding a structured repertoire of possible trajectories**, not merely a snapshot of past activity. This definition extends classical cell-assembly theory [10], [12] by emphasizing three properties:

1. **Dynamic expressivity**
   Engrams represent not only content (features, objects, episodes), but the *set of possible transitions* that content can undergo. Their structure supports perceptual recognition, recall, prediction and simulation through different activation pathways [18], [19].
2. **Contextual variability**
   The same engram may activate differently depending on neuromodulatory state, task demands or global network configuration [33], [34]. Thus, engrams behave as **programs** with context-sensitive execution, not as immutable memory traces.
3. **Multiscale embedding**
   Engrams exist across scales—from sensory microcircuits encoding local features to large hippocampal–cortical assemblies encoding episodes [11], [18]. Their hierarchical organization supports composition, generalization and abstraction [4], [5].

By conceptualizing engrams as dynamic neural programs, DIME captures the flexibility of neural representation and the ability of the same structural substrate to support perception, memory, imagination and planning. Without such a structure-centric unit, neural theories either oversimplify representation (as in point-neuron models) or fragment it across domain-specific modules.

### 3.2 Execution Threads: The Unit of Neural Process

While engrams define structural potential, cognition unfolds in time through coherent trajectories of neural activation. Empirical data consistently reveal such trajectories across diverse tasks—including sensory processing, sequential decision-making, motor planning and internal simulation [18], [19], [20].

DIME formalizes these trajectories as **execution threads**, defined as:

*Causally coherent, temporally extended paths of activation traversing engrame networks, driven by sensory input, prior state or internally generated dynamics.*

Execution threads unify phenomena often treated separately:

- **Perception**: feedforward–feedback sweeps forming stimulus-specific trajectories [26].
- **Memory recall**: hippocampal–cortical replay generating past-consistent paths [11].
- **Prediction and planning**: prefrontal–striatal sequences modeling potential future states [38].
- **Mind-wandering and spontaneous thought**: transitions in the default mode network [7], [25].

Threads provide the missing **process unit** that organizes neural dynamics into interpretable functional sequences. They also supply the substrate upon which marker systems act to modulate value, priority and learning.

Unlike static attractors or single-neuron spike trains, execution threads are **structured, adaptive and multimodal**, capable of merging sensory, mnemonic, motor and affective content into a unified temporal flow.

In the DIME architecture, cognitive control does not rely on a centralized executive component but emerges from the differentiated cooperation of the principal neuronal classes. Excitatory neuronal networks provide the structural substrate of engrams, defining the space of admissible activation states and the set of possible execution trajectories through recurrent connectivity and plasticity. Inhibitory neurons do not primarily encode representational content; instead, they implement temporal and competitive control over activation, regulating initiation, duration, frequency, and stability of execution threads while preventing nonspecific or unstable propagation. Neuromodulatory systems do not primarily define representational pathways, but adjust the global operating conditions of excitatory–inhibitory circuits, including excitability, activation thresholds, signal-to-noise ratio, and plasticity regimes. At larger spatial and temporal scales, volumetric neuromodulation and hormonal influences further shape the regional and physiological context in which DIME cycles operate. Together, these mechanisms enable stable, flexible, and context-sensitive neural programs without requiring a central controller, through distributed and emergent regulation.

### 3.3 Marker Systems: Value, Relevance and Control

Neural computation is not solely representational; it is fundamentally shaped by mechanisms that assign value, relevance and affective significance. Classical reinforcement learning models capture aspects of dopaminergic prediction error [13], [14], yet the brain employs a far richer set of modulatory processes—including serotonin, noradrenaline, acetylcholine, amygdala circuits and prefrontal control mechanisms [15], [21], [22].

DIME unifies these mechanisms under the notion of **marker systems**, defined as:

*Distributed neuromodulatory and limbic processes that regulate the amplification, suppression, stabilization and plastic modification of engrams and execution threads.*

Marker systems operate across spatial scales:

- **Local modulation**: synaptic gain adjustment, plasticity gating, attentional weighting [31], [14].
- **Regional modulation**: amygdalar tagging of emotional significance [21], [22].
- **Global modulation**: arousal, salience and goal-directed orientation mediated by brainstem–cortical loops [15], [25].

Markers determine:

- which percepts enter awareness,
- which memories are consolidated or suppressed,
- which decisions are favored under uncertainty,
- how prediction errors influence learning.

In this architecture, value is not an external label but an intrinsic neural property that **sculpts the space of threads and engrams in real time**. The absence of such a unified value mechanism is one of the major gaps in current cognitive theories.

### 3.4 Hyperengrams: Global Integration and Conscious Access

Neural evidence indicates that conscious experience requires large-scale integration across distributed cortical and subcortical networks [16], [7], [17], [26]. Such integration is dynamic, context-sensitive and modulated by both memory and affect [15], [25].

DIME introduces the concept of the **hyperengram**, defined as:

*A large-scale, temporally stable, value-marked network integrating identity, context, semantic structure, episodic history and goals.*

The hyperengram is not a static structure but a **continuously updated integration field**, emerging from interactions among active engrams, threads and marker systems. It has several core properties:

1. **Global availability**
   Hyperengrams recruit widespread networks (PFC, PCC, parietal cortex, hippocampus, thalamus) in a manner consistent with global broadcasting theories [16].
2. **Temporal continuity**
   They support the sense of a continuous self and perceptual coherence over seconds to minutes, distinguishing conscious from non-conscious processing [26], [24].
3. **Goal dependence**
   Marker systems modulate which components are included in the hyperengram, shaping attention, introspection and decision-making [15], [21].
4. **Narrative integration**
   Hyperengrams bind episodic traces, perceptual content and internal simulations into coherent sequences, forming the basis of conscious thought [25].

In DIME, consciousness emerges not from a specialized module, but from the **interaction** among engrams (structure), threads (process), markers (control) and hyperengrams (integration). This unified view bridges memory, perception, emotion and global awareness within a single operational architecture.

A fundamental architectural principle of DIME is the separation between an invariant operational mechanism and variable biological implementations. Unlike digital artificial systems, where the program is modifiable and the arithmetic-logic substrate is fixed, the biological brain operates through a stable and recurrent execution cycle—Detect–Integrate–Mark–Execute—that applies uniformly across levels and domains of processing. Cognitive variability does not arise from changes in this core algorithm, but from the continual reconfiguration of the neuronal structures on which it operates. In biological terms, the invariance of the DIME cycle corresponds to the stability of execution-thread dynamics, while functional diversity is implemented through the adaptive restructuring of engrams via inhibition, local and volumetric neuromodulation, and context-dependent plasticity. As a result, the same operational cycle can give rise to widely different interpretations, decisions, and behaviors without modification of the underlying algorithmic structure. In DIME, cognitive adaptation is achieved not by rewriting the program, but by reconfiguring the biological logic on which it runs, through distributed and emergent control.

The DIME framework thus defines cognition as the execution of a stable operational cycle over adaptive neural implementations, with value-dependent modulation providing selection, learning, and global coherence.

# 4. OPERATIONAL DISTINCTIVENESS OF THE DIME FRAMEWORK

The DIME architecture is not introduced as a replacement for existing theories, but as a proposal for an explicit operational unification across levels that are typically treated separately. Its distinctiveness lies not in the introduction of entirely new empirical entities, but in the formal integration of already documented mechanisms into a single recurrent computational cycle.

“Four aspects tentatively differentiate DIME at the architectural level:

### 4.1 An Explicit Process Unit: The Execution Thread

Many current frameworks describe neural dynamics in terms of attractor states, hierarchical message passing, or global broadcasting. However, they do not formalize a unified process unit that spans perception, memory, planning and spontaneous cognition.

DIME introduces the execution thread as an explicit temporal unit: a causally coherent trajectory traversing engram networks under marker modulation. This construct unifies perceptual sweeps, hippocampal replay, motor sequencing and internal simulation within a single formal category. In this sense, DIME treats cognition not as static state transitions, but as structured trajectory propagation.

### 4.2 Intrinsic Value as a Computational Parameter

In many computational models, value enters as an external reward signal or as a post hoc modulatory influence. In contrast, DIME treats marker systems as intrinsic control parameters governing:

- gain modulation,
- trajectory competition,
- plasticity gating,
- and stabilization of global states.

Value is therefore not an add-on to representation, but a structural dimension of the operational cycle itself. This positioning allows perception, memory consolidation, emotional salience and decision bias to be described as variations of the same underlying mechanism.

### 4.3 A Structural Account of Conscious Integration

Existing theories of consciousness emphasize broadcasting (Global Workspace), recurrence, or informational integration. DIME proposes a complementary structural perspective: conscious access corresponds to the formation of a hyperengram — a large-scale, temporally stabilized and marker-regulated integrative state.

This proposal does not deny broadcasting or recurrence; rather, it interprets them as mechanisms contributing to hyperengram stabilization. Consciousness is thus operationally defined as a particular configuration of the same engram–thread–marker architecture, scaled to global integration.

### 4.4 A Single Recurrent Cycle Across Scales

Perhaps the most distinctive feature of DIME is the claim that the Detect–Integrate–Mark–Execute loop operates at multiple nested scales:

- within local cortical microcircuits,
- across hippocampal–cortical interactions,
- within large-scale network reconfigurations,
- and in artificial DIME-inspired agents.

Rather than proposing separate algorithms for perception, memory and action, DIME formulates a single recurrent cycle whose instantiations differ only by scale, context and marker configuration. This provides a principled bridge between biological data and computational implementation.

**Short concluding paragraph (important)**

DIME therefore does not compete with predictive coding, engram theory or global workspace accounts at the level of isolated mechanisms. Instead, it proposes an explicit architectural closure in which these mechanisms are interpreted as components of a unified operational cycle. Whether this closure proves empirically adequate remains a matter for experimental testing, but it offers a coherent framework within which such testing can be systematically organized.

## 4.5 Testable Architectural Consequences

Beyond conceptual integration, the DIME framework entails architectural consequences that can, in principle, be empirically differentiated from existing models. These consequences arise not from isolated components, but from the coupling among engrams, execution threads and marker systems within a recurrent cycle.

**(1) Intermediate Marker-Competition Phase in Conscious Access**

DIME predicts that the transition from unconscious to conscious processing should involve a temporally distinct intermediate phase characterized by marker-mediated competition among execution threads before global stabilization.

Unlike models that predict either:

- a simple feedforward sweep followed by amplification, or
- an abrupt ignition event,

DIME implies a metastable interval during which competing trajectories are modulated by marker systems before a hyperengram achieves coherence.

This phase should be observable as:

- increased variability prior to large-scale synchronization,
- context-dependent modulation of trajectory stability,
- and task-dependent latency differences linked to valuation signals.

Failure to detect such an intermediate selection phase would weaken the architectural necessity of the marker–thread coupling.

**(2) Context-Dependent Trajectory Reconfiguration Within Stable Structural Engrams**

If engrams function as dynamic neural programs, the same structural network should be capable of generating different execution threads depending on marker configuration, without requiring slow synaptic restructuring.

DIME therefore predicts:

- rapid context-dependent trajectory divergence within the same neuronal population,
- modulation of sequence structure without major changes in connectivity,
- and task-dependent reconfiguration driven by neuromodulatory state.

If neuronal trajectories prove to be rigidly determined by connectivity alone and insensitive to rapid marker modulation, the DIME interpretation would require revision.

**(3) Scalable Recurrence Across Micro- and Macro-Levels**

Because DIME posits a single operational cycle across nested scales, it predicts structural similarity between:

- local cortical recurrence,
- hippocampal replay dynamics,

- large-scale network reconfigurations,
- and artificial DIME-based simulations.

This implies that similar temporal motifs (detect–integrate–mark–execute patterns) should be identifiable across spatial scales, differing primarily in complexity and integration scope.

If large-scale integration obeys principles fundamentally unrelated to local recurrent processing, the claim of scale invariance would be challenged.

**Final architectural clarification paragraph (very important)**

These predictions do not depend on the metaphysical status of any single component (engrams, threads, markers), but on the internal consistency of their interaction within a recurrent operational cycle. The value of the DIME framework therefore lies not in the novelty of its parts, many of which are grounded in established findings, but in the explicit architectural closure it proposes. Its empirical adequacy must ultimately be determined by whether such closure yields explanatory compression and experimentally discriminable consequences.

# 5. THE DIME ALGORITHM: FORMAL OUTLINE

The structural elements introduced in Section 3—engrams, execution threads, marker systems and hyperengrams—form the basis of an algorithmic cycle unifying perception, memory, prediction, valuation and action. This section formalizes the computational principles underlying this cycle and develops an abstract algorithmic outline suitable for implementation in artificial systems or for guiding neurobiological modeling.

## 5.1 Motivation for a Unified Algorithm

Neural activity displays consistent operational motifs across perceptual processing, episodic recall, predictive simulation and conscious deliberation [11], [12], [18], [19], [26]. Despite differences in content and context, these processes share four key computational demands:

1. **Pattern detection**: identifying relevant activity in sensory input or internal state [8], [35].
2. **Integration**: embedding new activity into ongoing internal trajectories [18], [20].
3. **Value assignment**: modulating activity according to goals, context and affect [14], [15], [21].
4. **Execution**: propagating selected trajectories to generate perception, recall, decision or overt action [38].

These operations recur across timescales and tasks, suggesting the existence of a **general-purpose neural algorithm**. DIME formalizes this idea by identifying Detect–Integrate–Mark–Execute as the primitive cycle coordinating neural computation across structures and scales. Rather than postulating distinct mechanisms for memory, attention, emotion or planning, DIME interprets these functions as **specializations of a single cycle** acting on different engrams, threads and marker configurations.

## 5.2 Formal Components

The algorithm relies on four functional components:

**(1) Engram space $\mathcal{E}$**

A set of engrams

$$\mathcal{E} = \{E_1, E_2, \ldots, E_N\}$$

where each engram $E_i$ is defined by:

- a structural substrate (connectivity matrix),
- a set of admissible activation patterns,
- a family of transitions between patterns.

**(2) Thread space $\mathcal{T}$**

A set of execution threads

$$\mathcal{T} = \{\tau_1, \tau_2, \ldots\}$$

where each thread $\tau$is a temporally ordered sequence of state transitions across engrams.

**(3) Marker field M(t)**

A vector-valued function representing the instantaneous state of neuromodulatory and limbic systems:

$$M(t) = \sum_{i=1}^{k} w_i \, S_i(t)$$

where $S_i(t)$is the activity of the i-th marker source (e.g., dopamine, amygdala, noradrenaline) and $w_i$are context-dependent weights [14], [15], [21].

**(4) Hyperengram H(t)**

A large-scale integrative state defined as:

$$H(t) = \Phi(\mathcal{T}(t), M(t))$$

where $\Phi$ is a functional integrating active threads and marker state into a coherent global representational field [16], [7], [24].

## 5.3 Core Computational Cycle

At time t, the DIME cycle proceeds as follows:

**Step 1. Detect**

Incoming sensory flux $S(t)$or internal signals are matched against engram patterns:

$$D(t) = \text{match } (S(t), \mathcal{E})$$

This step includes pattern completion, novelty detection and ambiguity resolution [10], [35], [18].

**Step 2. Integrate**

Detected engrams update the currently active execution threads:

$$\mathcal{T}(t+1) = \text{integrate } (D(t), \mathcal{T}(t), H(t))$$

Integration may merge, split or extend threads, depending on prior context and global state [11], [19], [20].

**Step 3. Mark**

Marker systems assign value, salience and affective weight to active engrams and threads:

$$W(t+1) = f(M(t), \mathcal{T}(t+1), D(t))$$

This modulates amplification, suppression and plasticity [14], [21], [22].

**Step 4. Execute**

Selected threads propagate, producing perceptual binding, recall, prediction or motor output:

$$O(t+1) = \text{execute } (\mathcal{T}(t+1), W(t+1))$$

Execution may be covert (internal simulation) or overt (behaviour).

These four steps form a recurrent computation, continuously updating both local engrams and the global hyperengram.

### 5.4 Multi-Scale Operation

The DIME cycle operates simultaneously across:

- **millisecond-level** (spiking, cortical microcircuits) [32],
- **tens–hundreds of milliseconds** (perceptual integration, recurrent processing) [26],
- **seconds** (working memory, conscious access) [16],
- **minutes–hours** (episodic consolidation, planning) [11],
- **days–years** (long-term knowledge formation).

Each timescale interacts via nested loops: fast cycles shape slow cycles and vice versa. This avoids the fragmentation inherent in many existing theories where different timescales require distinct models.

### 5.5 Match, Mismatch and Predictive Divergence

Although DIME incorporates elements of predictive processing, it differs in key respects from classical predictive coding [8], [9].

Engram detection yields three possible outcomes:

- **Match:** activation falls within expected engram patterns.
- **Partial match:** triggering pattern completion or ambiguity resolution.
- **Mismatch:** indicating novelty, prompting thread divergence and high marker reactivity [35], [18].

Predictive divergence does not merely generate error-minimization signals; it **reconfigures thread dynamics** and may prompt new engram formation or updating, under strong marker influence [14], [37].

### 5.6 Marker-Driven Modulation

The marker field M(t):

- scales thread gain,
- gates plasticity,
- increases or decreases exploration,
- biases recall vs. prediction,
- shapes decision thresholds [14], [15], [21].

For example:

- High dopamine biases learning and forward exploration [14];
- Low serotonin increases impulsivity and noise tolerance;
- Amygdalar activation tags threads with emotional relevance [21], [22].

DIME thus integrates theories of reinforcement learning, affective modulation and executive control into a unified computational mechanism.

### 5.7 Learning and Plasticity in DIME

Learning arises from coordinated changes in:

- **engram connectivity** (Hebbian, STDP, neuromodulator-gated plasticity [31], [14]);
- **thread selection** (reinforcement via marker states);
- **hyperengram configuration** (changes in global integration patterns [16], [24]).

Unlike static ANN training, learning in DIME is:

- **online**,
- **state-dependent**,

- **marker-gated**,
- **multi-scale**,
- **structurally flexible**.

  The system updates continuously rather than via epoch-based weight optimization.

### 5.8 Algorithmic Pseudocode (High-Level)

```
Initialize Engram Space 𝓔
Initialize Thread Set 𝓣 = {}
Initialize Marker Field M
Initialize Hyperengram H

For each timestep t:

    # 1. Detect
    D = match(S(t), 𝓔)

    # 2. Integrate
    𝓣 = integrate(D, 𝓣, H)

    # 3. Mark
    W = marker_update(M, 𝓣, D)

    # 4. Execute
    O = execute(𝓣, W)

    # Hyperengram update
    H = update_hyperengram(𝓣, M)

Return O
```

This is not a mechanistic model of neural dynamics, but an abstract algorithm capturing the computational essence of the Detect–Integrate–Mark–Execute cycle.

The pseudocode abstracts away from biophysical implementation details and should not be interpreted as a literal neural simulation.

## 6. NEUROBIOLOGICAL GROUNDING OF DIME

The DIME architecture is constructed not as an abstract computational metaphor but as a synthesis of convergent findings across systems neuroscience, cognitive neuroscience and cellular neurobiology. Each component — engrams, execution threads, marker systems and hyperengrams — corresponds to empirically documented mechanisms operating across structural and temporal scales. This section reviews key evidence supporting the biological plausibility of the DIME framework.

### 6.1 Evidence for Engrams as Dynamic Neural Programs

Neuroscience has produced robust evidence that memory, perception and learned representations are encoded not in individual neurons but in **distributed neural ensembles** whose activation patterns can be reinstantiated across time [10], [11], [12]. These assemblies exhibit several properties fundamental to the DIME definition of engrams:

**1. Pattern completion and reconstruction**

Hippocampal and cortical engrams can be reactivated by partial cues, reconstructing full representational states [10], [35]. This demonstrates that engrams contain structured activation repertoires rather than fixed-point states.

**2. State-dependent reconfiguration**

The same engram can exhibit multiple activation profiles depending on neuromodulatory state or behavioural context [33], [21], indicating that engrams function as **programmable substrates** rather than static memory traces.

**3. Multiscale hierarchical embedding**

From simple sensory feature maps [32], [34] to complex episodic networks in hippocampus–cortex loops [11], [18], engrams exist at multiple levels of abstraction, mirroring the hierarchical representational structure required by DIME.

**4. Reconsolidation and updating**

Engrams undergo dynamic modification during recall and reconsolidation [37], showing that they encode adaptable procedures rather than immutable records.

Taken together, these findings strongly support the **program-like** interpretation of engrams central to DIME: distributed, recurrent networks capable of generating multiple state trajectories, modulated by context and history.

## 6.2 Evidence for Execution Threads as Spatiotemporal Trajectories

Neural activity is not a static mapping between stimuli and responses but unfolds as **structured temporal trajectories** linking distributed assemblies [18], [19], [20]. Empirical evidence for execution threads includes:

**1. Sequential activation in hippocampal–cortical loops**

Replay and preplay phenomena demonstrate temporally ordered activation across ensembles during recall, simulation and planning [11], [38].

**2. Dynamic trajectories during perceptual integration**

Recurrent processing in visual cortex produces rapid forward–backward sweeps that integrate sensory fragments into coherent percepts [26].

**3. Prefrontal trajectory-based planning**

Neurons in prefrontal cortex encode not only current state but extended future-directed sequences [35], consistent with thread-like propagation.

**4. Intrinsic trajectories in default mode dynamics**

Resting-state brain activity displays continuous transitions through a structured manifold of network states [7], [25].

Across tasks, species and modalities, the brain consistently expresses **time-evolving, context-sensitive activation paths**, matching the definition of execution threads in DIME.

## 6.3 Evidence for Marker Systems as Value and Control Mechanisms

Value assignment, salience modulation and behavioural control rely on **distributed neuromodulatory and limbic systems**, whose functions align precisely with the DIME concept of marker systems.

**1. Dopaminergic modulation of learning and exploration**

Dopamine encodes prediction error, reward expectation and motivational value, modulating plasticity and behavioural policy [13], [14].

**2. Amygdalar tagging of emotional relevance**

Amygdala circuits bind affective significance to perceptual and mnemonic content [21], [22], shaping the likelihood of consolidation and recall.

**3. Noradrenergic regulation of arousal and uncertainty**

Locus coeruleus activity modulates exploration–exploitation tradeoffs and sensory gain control [15].

**4. Cholinergic modulation of attention and plasticity**

Acetylcholine enhances signal-to-noise ratio in sensory processing and gates learning in cortical circuits [31], [14].

**5. Serotonergic modulation of patience, impulsivity and confidence**

Serotonin shapes behavioural persistence, valuation of delayed outcomes and tolerance of uncertainty.

These systems operate not in isolation but as an **interacting control layer** that regulates which engrams and threads are amplified, stabilized or suppressed. This supports the DIME view that value is not a separate module but a global computational parameter shaping cognition.

## 6.4 Evidence for Hyperengrams and Large-Scale Integration

Large-scale integration is a hallmark of conscious access and high-level cognition [16], [17], [24]. Several empirical findings support the concept of hyperengrams:

**1. Global broadcasting and fronto–parietal recruitment**

Conscious perception correlates with widespread activation across prefrontal, parietal and cingulate networks [16], [17].

**2. Default mode network as a substrate for narrative integration**

DMN dynamics integrate autobiographical memory, semantic knowledge, future simulation and self-referential processing [7], [25].

**3. Hippocampal–cortical indexing**

The hippocampus binds distributed cortical representations into coherent episodes, a core mechanism for the formation of hyperengrams [11], [35].

**4. Network-level coherence and synchrony**

Gamma, theta and alpha synchrony across distant regions predict the stability and vividness of conscious states [18], [24].

These findings align with the DIME definition of hyperengrams as **value-marked, large-scale, temporally extended integrative states** linking identity, context, history and goals.

## 6.5 Convergence: DIME as a Synthesis of Known Mechanisms

Each DIME component corresponds to an independently validated neural phenomenon:

| DIME Component | Neurobiological Evidence | Sources |
|---|---|---|
| **Engrams** | Distributed assemblies, pattern completion, reconsolidation | [10], [11], [12], [35], [37] |
| **Execution Threads** | Replay, planning trajectories, recurrent perceptual sweeps | [11], [19], [26], [20], [38] |
| **Marker Systems** | Dopamine RL, amygdalar tagging, cholinergic and noradrenergic modulation | [13], [14], [15], [21], [22] |
| **Hyperengrams** | Global broadcasting, DMN integration, network-level synchrony | [16], [7], [17], [24], [25] |

DIME therefore synthesizes empirically supported mechanisms into a **coherent computational architecture**, bridging structural, dynamical, modulatory and integrative levels of brain organization.

## 6.6 Conceptual Diagrams of the DIME Architecture

To facilitate understanding of the interactions among structural, dynamical, modulatory and integrative components, this section presents schematic diagrams illustrating the core conceptual elements of the DIME architecture. The diagrams are conceptual in nature and do not specify anatomical details; instead, they capture the computational topology and flow of influence among engrams, execution threads, marker systems and hyperengrams. More detailed graphical representations are provided in the companion monograph [1].

**Figure 1. Multi-layer Schematic of the DIME Architecture**

```
                  +-----------------------------------+
                  |            HYPERENGRAM            |
                  |    (Global Integration Layer)     |
                  +-----------------------------------+
                              ▲             ▲
                              |             |
                    Marker Feedback     Global Context
                              |             |
+-----------------------------+-------------+-----------------------------+
|                         MARKER SYSTEMS (M)                              |
|   Dopamine | Serotonin | Noradrenaline | Acetylcholine | Amygdala | PFC     |
+-----------------------------+-------------+-----------------------------+
                              |             |
                              | Modulation  |
                              ▼             ▼
                    +-----------------------------+
                    |      EXECUTION THREADS      |
                    |  (Dynamic Spatiotemporal    |
                    |        Trajectories)        |
                    +-----------------------------+
                              ▲             ▲
                              | Thread Flow |
                              ▼             ▼
+-----------------------------+-------------+-----------------------------+
|                         ENGRAM SPACE (E)                                |
|   Distributed recurrent networks encoding structured activation patterns    |
+-------------------------------------------------------------------------+
```

**Description:**

This diagram organizes the architecture into four interacting layers:

1. **Structural Layer (Engrams — representational substrate)**
   Distributed recurrent networks encoding representational potential.
2. **Dynamical Layer (Execution Threads — process-level trajectories)**
   Directed spatiotemporal trajectories connecting active engrams.

3. **Control Layer (Marker Systems — value and gain modulation)**
   Neuromodulatory and limbic influences modulating thread gain, plasticity and selection.
4. **Integrative Layer (Hyperengrams — global integration states)**
   Large-scale integration of identity, context, goals and episodic history.
   **Conceptual layout (textual schematic):**

**Interpretation:**

Cognition emerges from vertical interactions (structure → process → control → integration) and horizontal dynamics (feedback loops across layers). Threads traverse engrams, marker systems reshape thread trajectories, and hyperengrams integrate system-level states. **The same multi-layer organization operates across local, intermediate and global scales, defining a single operational architecture rather than separate cognitive modules.** These interactions do not form a command hierarchy, but a recurrent functional loop underlying the Detect–Integrate–Mark–Execute cycle.

**Figure 2. Engrams as Dynamic Neural Programs**

**Description:**

This schematic illustrates that an engram is not a static cluster of neurons, but a structured state-space substrate supporting multiple admissible activation trajectories.

**Conceptual layout:**

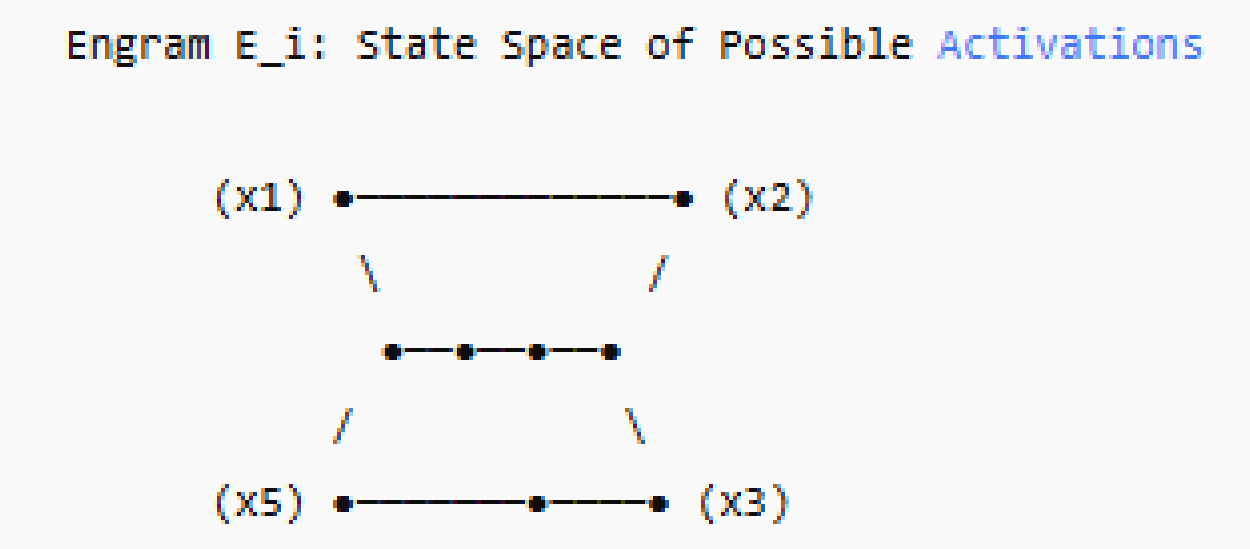

**Interpretation:**

An engram encodes a repertoire of possible state transitions rather than a fixed memory trace. These trajectories are **selectable, context-dependent and modulatable**, allowing the same engram to participate in perception, memory retrieval, prediction and internal simulation. This captures the concept of engrams as dynamic neural programs, capable of pattern completion, generalization and flexible reconfiguration under marker influence, rather than as static attractor states [10], [11], [12], [37].

**Figure 3. Execution Threads as Spatiotemporal Trajectories**

**Description:**

This diagram illustrates an execution thread as a temporally extended trajectory traversing multiple engram networks.

**Conceptual layout:**

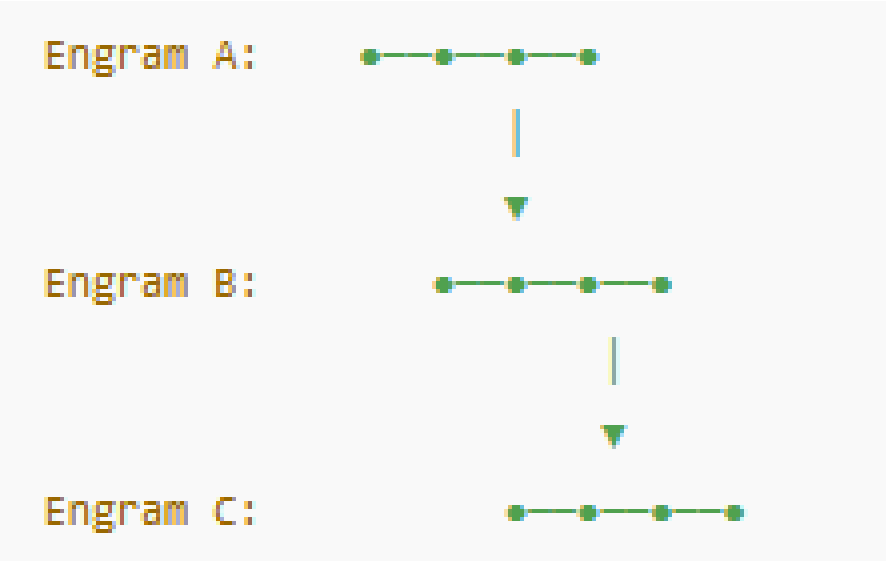

**Interpretation:**

Each arrow represents a context-sensitive transition shaped by prior activation history, predictive structure and marker state. Execution threads constitute the **fundamental unit of neural process in DIME**, forming causally coherent activation trajectories rather than isolated state transitions. Multiple threads may coexist and compete in parallel, with their continuation, branching or suppression regulated by marker systems. Through this mechanism, execution threads integrate past experience, ongoing perception and projected future states into a unified temporal process [18], [19], [20], [38].

**Figure 4. Marker Systems as Global and Local Modulators**

**Description:**

This schematic illustrates marker systems as distributed control fields influencing neural processing across multiple spatial and temporal scales.

**Conceptual layout:**

```
    +-------------------------------------------+
    |              MARKER FIELD M(t)             |
    |  [DA] [5-HT] [NE] [ACh] [Amygdala] [PFC]   |
    +-------------------------------------------+

               ↓ Gain / Value / Plasticity

+----------+    +-----------+     +-----------+
| Engram 1 | <-- | Thread X | -->  | Engram 2  |
+----------+    +-----------+     +-----------+
```

**Interpretation:**

Marker systems modulate core computational parameters rather than representational content, including:

• excitability and signal-to-noise ratio (noradrenaline, acetylcholine),
• learning rate and trajectory reinforcement (dopamine),
• affective tagging and relevance assignment (amygdala),
• decision flexibility and persistence (serotonin).

Crucially, marker systems **do not select or store content**. Instead, they reshape the conditions under which execution threads propagate, stabilize or are suppressed, thereby biasing the space of possible neural trajectories without invoking a centralized controller. This distributed modulation enables value-based selection, learning and adaptive behavior to emerge intrinsically from system dynamics [13], [14], [15], [21], [22].

**Figure 5. Hyperengrams as Large-Scale Integrative Networks**

**Description:**

This schematic provides a high-level representation of how interactions among engrams, execution threads and marker systems give rise to a large-scale integrative state.

**Conceptual layout:**

```
Engram A     Engram B     Engram C
    \            |            /
     \           |           /
      Thread 1   Thread 2   Thread 3
            \      |      /
             \     |     /
              +----------------+
              |  HYPERENGRAM   |
              | (Global State) |
              +----------------+
```

**Interpretation:**

Hyperengrams correspond to value-marked, temporally stabilized integration fields emerging from the coordinated activity of multiple execution threads across distributed engram networks. Rather than constituting a dedicated module or instantaneous broadcast mechanism, hyperengrams reflect the sustained convergence of structure, process and control over time. Through this stabilization, distributed perceptual, mnemonic, affective and goal-related content becomes globally available, supporting conscious access, narrative continuity and coherent decision-making [16], [7], [17], [24], [25].

**Figure 6. Full DIME Algorithm Cycle**

**Description:**

This diagram presents a recurrent flow illustrating the Detect–Integrate–Mark–Execute (DIME) operational cycle.

**Conceptual layout:**

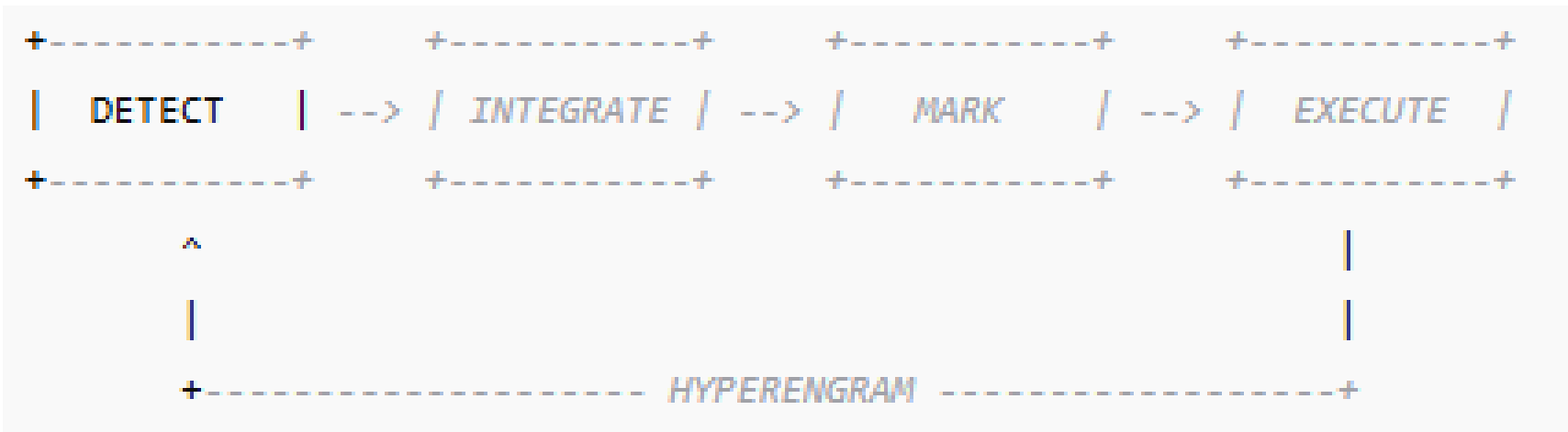

**Interpretation:**

The diagram shows how all DIME components interact algorithmically over time within a single recurrent loop. At each iteration, the cycle updates local dynamics (engrams and execution threads) and the global integrative state (hyperengram), under continuous modulation from marker systems. The same operational cycle governs perception, memory retrieval, internal simulation, decision-making and action, across multiple spatial and temporal scales. Importantly, the DIME cycle does not define a linear pipeline with privileged entry or exit points, but a continuous, self-updating process that constitutes the core computational mechanism of the architecture.

Diagram 7 — DIME-Based AI System Architecture

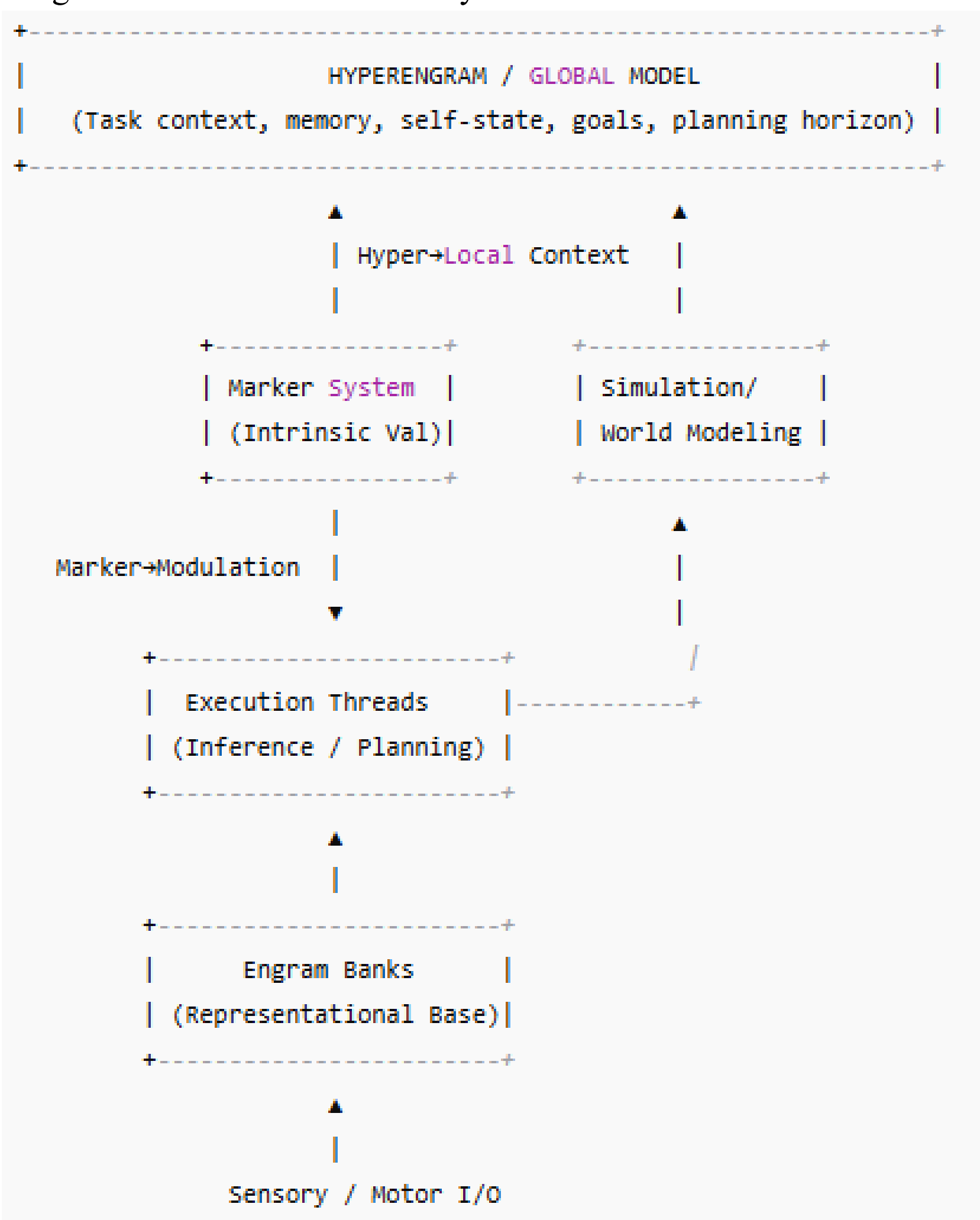

Diagram 8 — DIME-Based Robotic Control Loop

```
 Sensors
   ↓
[Detect] → perceptual engrame activation
   ↓
[Integrate] → thread update (planning + memory)
   ↓
[Mark] → motivation, urgency, value, uncertainty
   ↓
[Execute] → motor commands + predictive simulation
   ↓
 Hyperengram Update → global coherence for next cycle
```

### 6.7 Summary

The diagrams in this section visualize the computational architecture underlying DIME:

- **Figure 1:** Global multi-layer architecture
- **Figure 2:** Engrams as dynamic programs
- **Figure 3:** Execution threads as trajectories
- **Figure 4:** Marker systems as modulators
- **Figure 5:** Hyperengrams as global integrators
- **Figure 6:** The DIME algorithmic loop

These diagrams collectively illustrate how structure, dynamics, value and integration interact to support cognition and consciousness. They will serve as the basis for graphical figures in the final arXiv version and are fully elaborated in the monograph [1].

## 7. DIME IN RELATION TO EXISTING THEORIES

DIME does not reject existing neuroscientific and computational frameworks; instead, it **integrates their strongest explanatory elements** while addressing several structural and dynamical gaps identified in Sections 2–5. This section compares DIME with influential theoretical traditions, highlighting both convergences and divergences.

### 7.1 Relation to Predictive Coding (PC)

Predictive coding proposes that the brain maintains hierarchical generative models that minimize prediction error through reciprocal message passing between cortical layers [8], [9].

**Convergences:**

- Both PC and DIME emphasize **top-down contextual influences** on local processing.
- Both treat the brain as an **active inference machine**, continuously integrating prior expectations with incoming evidence.
- DIME's **Detect** step partially corresponds to pattern matching and prediction-error evaluation in PC.

**Key Divergences:**

1. **Unit of computation**
    - PC operates on single-neuron or population-level error units.
    - DIME operates on **engrams and execution threads**, which are *multiscale*, *representational* and *temporally extended* units.

2. **Nature of inference**
    - PC proposes continuous error minimization.
    - DIME allows **non-minimizing divergence** (e.g., novelty-triggered exploration, thread re-routing), guided by marker systems [14], [15].
3. **Role of value**
    - PC lacks an explicit representation of motivational or affective value.
    - DIME incorporates **marker systems** as first-class computational variables shaping inference, learning and selection.
4. **Global integration**
    - PC is primarily local and hierarchical.
    - DIME includes **hyperengrams** supporting narrative continuity and conscious access.

Thus, PC is partially subsumed by the Detect–Integrate steps of DIME but does not account for thread dynamics, marker-driven modulation or hyperengram integration.

## 7.2 Relation to Active Inference (AIx) and the Free-Energy Principle

Active inference formalizes perception and action as processes minimizing variational free energy by adjusting beliefs and motor policies [8].

**Convergences:**

- Both frameworks treat cognition as a **closed-loop** between perception and action.
- Both emphasize probabilistic inference shaped by prior expectations.
- DIME's **Execute** step is consistent with the active selection of policies.

**Key Divergences:**

1. **State representation**
    - Active inference uses latent-variable generative models.
    - DIME uses **engrams**, biological substrates encoding actual possible trajectories, not idealized latent variables.
2. **Temporal unfolding**
    - Active inference models inference as optimization.
    - DIME models it as **trajectory evolution** of execution threads, not as gradient descent.
3. **Modulatory control**
    - Active inference treats value as precision weighting.
    - DIME incorporates **distinct marker systems** (dopamine, NE, ACh, amygdala), with empirically grounded neurobiology [13], [14], [21].
4. **Conscious integration**
    - Active inference has no explicit mechanism for subjective unity.
    - DIME introduces **hyperengrams** as the basis for global workspace-like integration.

DIME can be viewed as a biologically grounded, multi-scale reformulation that includes active inference as one special case of its general algorithmic cycle.

## 7.3 Relation to Memory Engram Theories

Engram theory focuses on the physical substrates of memory in the form of reactivatable neural ensembles [10], [11], [37].

**Convergences:**

- Both DIME and classical engram theory posit distributed assemblies as fundamental memory units.
- Both account for pattern completion, reconsolidation and context sensitivity.

**Key Divergences:**

1. **Functional scope**
   - Classical engram theory explains **memory encoding and retrieval**.
   - DIME expands engrams to **general representational units** participating in perception, planning, prediction, and conscious thought.
2. **Dynamics**
   - Engram theory describes reactivation but not its integration into extended temporal sequences.
   - DIME introduces **execution threads** as the dynamical complement.
3. **Control**
   - Engram theory lacks a complete account of neuromodulatory value assignment.
   - DIME embeds engrams within **marker-modulated selection mechanisms**.

Thus, DIME generalizes engrams from memory-specific traces to multi-role dynamic neural programs.

### 7.4 Relation to Attractor Network Models

Attractor networks model memory, categorization and decision-making via convergence to stable activation patterns [12].

**Convergences:**

- Both DIME and attractor frameworks emphasize recurrent connectivity and multistability.
- Both recognize that neural states evolve through structured state spaces.

**Key Divergences:**

1. **Attractor dynamics vs. trajectory dynamics**
   - Attractor networks converge to **static fixed points**.
   - DIME emphasizes **open-ended trajectories** (threads) that may traverse, merge, diverge or avoid attractors.
2. **Scope**
   - Attractor models usually describe one function (memory, decision).
   - DIME spans **perception, action, memory, emotion and consciousness**.
3. **Global integration**
   - Attractor models have no global narrative mechanism.
   - DIME introduces **hyperengrams** for multi-level coherence.

Thus, DIME preserves local attractor-like mechanisms but situates them within a broader architecture capable of representing extended cognitive sequences.

### 7.5 Relation to Global Workspace Theory (GWT)

Global Workspace Theory proposes that consciousness arises when information becomes globally available through widespread cortical broadcasting [16], [17].

**Convergences:**

- DIME’s **hyperengram** is consistent with global broadcasting phenomena.
- Both assume that conscious access requires **integration across distant regions**.

- Both explain working memory as a stable, sustained global state.

**Key Divergences:**

1. **Mechanism of integration**
   - GWT describes what becomes conscious but not how dynamic structures generate it.
   - DIME provides **algorithmic mechanisms** (thread dynamics + modulation + hyperengram).
2. **Temporal structure**
   - GWT lacks detailed temporal sequencing rules.
   - DIME models subjective continuity via **thread-to-hyperengram updates**.
3. **Value and emotion**
   - GWT underrepresents affective modulation.
   - DIME explicitly incorporates **marker systems** into conscious integration.

Thus, DIME provides a mechanistic substrate for GWT-like phenomena.

## 7.6 Relation to Reinforcement Learning and Decision Models

Computational reinforcement learning (RL) models focus on reward prediction error, policy learning and value-based decision processes [14], [38].

**Convergences:**

- Both RL and DIME use value signals to shape behaviour.
- Dopamine prediction error appears in DIME as part of the **marker field M(t)**.

**Key Divergences:**

1. **Representational substrate**
   - RL uses discrete states or vector embeddings.
   - DIME uses **engrams**, biologically grounded representational units.
2. **Temporal control**
   - RL policies are discrete action selectors.
   - In DIME, **threads generate behavior** and can span multiple actions.
3. **Emotion, uncertainty and context**
   - RL lacks serotonin, NE, ACh, amygdalar and PFC modulation.
   - DIME integrates these as **marker components** shaping learning, exploration and decision thresholds [15], [21].
4. **Global integration**
   - RL does not address conscious-level integration.
   - DIME integrates decision-making into the hyperengram layer.

Thus, RL describes a subset of marker-modulated processes that DIME generalizes to the full cognitive cycle.

## 7.7 Relation to Dynamical Systems Models

Neural activity is often modeled as movement through a high-dimensional state space governed by differential equations [19], [20].

**Convergences:**

- Both DIME and dynamical systems approaches emphasize continuous trajectories.
- Both reject static representations.

**Key Divergences:**

1. **Level of description**
    - Dynamical models typically describe the brain at the **population vector** level.
    - DIME's execution threads are **functional trajectories** through representational structures.
2. **Control**
    - Dynamical models rarely include neuromodulatory markers.
    - DIME explicitly models **marker-gated trajectory selection**.
3. **Global narrative**
    - Dynamical systems models lack a construct analogous to the hyperengram.
    - DIME adds a global integration variable necessary for subjective continuity.

DIME thus incorporates dynamical systems behavior but embeds it in a multi-layer computational architecture.

### 7.8 Summary: What DIME Adds

Across all comparisons, several unifying themes emerge:

**What is shared with existing theories**

- recurrent processing,
- hierarchical integration,
- probabilistic inference,
- distributed representation,
- dynamic trajectories,
- global broadcasting.

**What DIME explicitly integrates beyond existing frameworks:**

1. **A unified structural unit**: engrams as dynamic neural programs.
2. **A unified process unit**: execution threads as spatiotemporal carriers of cognition.
3. **A unified control mechanism**: marker systems integrating emotion, motivation and cognitive modulation.
4. **A unified integrative substrate**: hyperengrams enabling conscious access.
5. **A unified algorithmic cycle**: Detect–Integrate–Mark–Execute across all cognitive domains.

Thus, DIME can be interpreted as a synthetic meta-architecture that attempts to integrate and reinterpret the strengths of existing frameworks and approaches.

## 8. IMPLICATIONS FOR ARTIFICIAL INTELLIGENCE AND ROBOTICS

This section discusses architectural implications for AI and robotics, not benchmarked implementations.

Although DIME is grounded in neurobiology, its architectural principles are sufficiently abstract and computationally well-formed to inspire the design of artificial cognitive systems. This section outlines how DIME can inform next-generation AI models and robotic architectures, highlighting advantages over contemporary deep learning, reinforcement learning and hybrid neuro-symbolic approaches.

## 8.1 Limitations of Current AI Architectures

Modern AI systems exhibit remarkable performance across narrow tasks but face structural limitations:

**1. Lack of unified representation and process**

Deep learning models use fixed architectures with static computation graphs, unlike the flexible engram–thread structure in the brain [27], [28].

**2. Absence of intrinsic value mechanisms**

Reward signals in RL are externally defined; AI systems lack intrinsic modulatory systems comparable to dopamine, serotonin or amygdala-based affect [14], [21].

**3. No internal narrative or global integrative state**

Transformers maintain context via attention windows but do not produce a persistent global integrator analogous to the hyperengram [32].

**4. Temporal brittleness**

AI models operate in discrete steps with limited long-range temporal coherence; real cognition is continuous and trajectory-based [29].

**5. Non-embodied inference**

Most AI models ignore internal physiological states (marker fields), yet these states are essential for adaptive behaviour in biological agents [15].

These limitations motivate research into architectures that integrate representation, value and control into a single coherent mechanism.

## 8.2 Engrams as a Basis for Adaptive AI Representations

DIME proposes engrams as **dynamic representational programs**, offering several advantages for AI:

**(a) Multi-level abstraction**

Engrams can encode both fine-grained features and high-level conceptual structures, similar to multi-layer deep networks but **not restricted to fixed layers**.

**(b) Context-sensitive reconfiguration**

Instead of static embeddings, an engram's activation repertoire can change depending on internal marker states or task demands [33], [37].

**(c) Unified representation for memory, perception and planning**

In contrast to architectures separating memory modules, feature extractors and policy networks, DIME uses **engrams for all representational roles**, reducing fragmentation.

**(d) Efficient reuse**

Engrams can participate in multiple tasks (perception, imagination, recall), enabling **dynamic compositionality** absent in most current AI models.

These properties suggest that engram-inspired architectures may outperform traditional embeddings, VAEs, or transformer token representations in terms of generalization and robustness.

## 8.3 Execution Threads as Trajectory-Based Computation in AI

Execution threads provide a **general mechanism for temporal reasoning, planning and internal simulation**.

**Advantages for AI:**

1. **Trajectory-level computation**

Unlike static feedforward inference, threads allow models to reason over extended time horizons, akin to neural ODEs or recurrent state-space models.

2. **Unified mechanism for perception, planning and imagination**
   Threads correspond to multi-step rollouts, bridging model-based RL, generative simulation and working memory.
3. **Contextual branching**
   Threads can branch or merge depending on marker influence, resembling tree search or multi-hypothesis simulation but with biological grounding.
4. **Flexible horizon control**
   Robust long-term planning emerges from thread continuation rather than from fixed-horizon optimization.

Threads therefore provide a biologically motivated alternative to transformer recurrence, recurrent state-space models, or Monte Carlo planning.

## 8.4 Marker Systems as Intrinsic Value Mechanisms for AI

Current reinforcement learning assigns reward externally; DIME introduces **value as an internal modulatory field**.

**Potential benefits for AI:**

- **Adaptive learning rates** modulated by intrinsic marker signals (dopamine analogs).
- **Dynamic exploration/exploitation balance** influenced by artificial noradrenaline-like signals.
- **Affective tagging** of experiences using amygdala-inspired mechanisms to guide memory retention.
- **Contextual weighting** shaping attention, prediction, and planning.

This could solve long-standing challenges:

- fragile reward shaping,
- instability in RL training,
- delayed credit assignment,
- lack of transferability across tasks.

Marker-inspired AI systems would be capable of **autonomous motivation**, internal prioritization, and state-dependent learning—abilities absent in conventional RL.

## 8.5 Hyperengrams as Global States for Synthetic Consciousness and System-Level Coherence

Hyperengrams offer an architectural solution to the **global integration problem** in AI.

**Key advantages:**

1. **Persistent global memory**
   Functions like a working-memory–plus narrative state, superior to transformer context windows [29].
2. **Unified identity and goal representation**
   Facilitates long-term coherence and self-consistency across tasks.
3. **Multi-modal integration**
   Hyperengrams bridge vision, motor plans, semantic memory and episodic recall.
4. **Basis for introspection and metacognition**

With hyperengrams, a model can evaluate its own internal state in a structured way, enabling meta-reasoning.

Hyperengrams therefore provide a biologically grounded analogue to **world models**, **global latent spaces**, and **self-consistency modules**.

## 8.6 Towards DIME-Based Artificial Agents

A DIME-inspired artificial agent would include:

- **Engram-like representational modules** with dynamic activation topologies.
- **Thread-based inference cycles** implementing multi-step reasoning, prediction and action selection.
- **Marker fields** controlling learning, attention and behavioural strategies.
- **A hyperengram-like global workspace** aligning local computations with system-level goals.

Such agents would exhibit:

- improved **generalization**,
- **adaptive exploration**,
- **hierarchical planning**,
- **transfer learning**,
- **robustness to noise and ambiguity**,
- and potentially early forms of **synthetic self-awareness,**
- if successfully implemented.

These properties align with ongoing work on neuro-symbolic integration, model-based RL, autonomous robotics and hybrid deep learning–dynamical systems models.

## 8.7 Robotics: Embodied Implementation of DIME

Robots provide an ideal testbed for DIME due to their need for:

- real-time perception,
- continuous control,
- hierarchical planning,
- adaptive learning,
- and integration of sensorimotor, affective and contextual signals.

**Potential robotic advantages of DIME:**

1. **Trajectory-based control**
   Execution threads map naturally to hierarchical motor control and sensorimotor loops.
2. **Embodied marker fields**
   Robots can implement synthetic analogs of physiological states (battery, temperature, wear), modulating behaviour like biological markers.
3. **Resilience and fault-tolerance**
   Hyperengram integration supports graceful degradation under partial failure.
4. **Unified perception–action loop**
   Detect–Integrate–Mark–Execute becomes a closed-loop controller with adaptive internal state.

Such robots would differ fundamentally from current systems based on precomputed policies or static networks.

### 8.8 Summary: DIME as a Bridge Between Neuroscience and AI

DIME offers a biologically grounded path toward **General Cognitive Architectures (GCA)** by unifying:

- neural structure (engrams),
- dynamics (threads),
- value (markers),
- integration (hyperengrams),
- and computation (DIME cycle).

This synthesis addresses the shortcomings of current AI systems and aligns with emerging trends toward **agentic models**, **world-simulating architectures**, **neural ODEs**, and **reinforcement learning with intrinsic motivation**.

DIME may offer a conceptual orientation for the next generation of cognitive AI and autonomous agents.

## 9. DISCUSSION AND OUTLOOK

The DIME architecture proposes a unified framework for understanding how the brain constructs representations, sequences internal activity, assigns value, and integrates information into coherent conscious experience. By grounding each computational component in neurobiological evidence, DIME aims to bridge the gap between descriptive neuroscience and mechanistic cognitive theory.

This section outlines the broader implications, theoretical strengths, current limitations, and future research directions associated with DIME.

### 9.1 Summary of Contributions

DIME provides a multi-layered synthesis built around four core innovations:

**1. Engrams as dynamic neural programs**
Instead of treating engrams as static memory traces, DIME interprets them as **structured dynamical substrates** that support perception, memory, abstraction, imagination and planning [10], [11], [37].

**2. Execution threads as the fundamental unit of neural process**
DIME reframes cognitive operations as **spatiotemporal trajectories**, integrating phenomena traditionally attributed to separate mechanisms: perception, recall, prediction, decision-making and motor planning [18], [19], [20], [38].

**3. Marker systems as intrinsic control and valuation**
Motivation, affect, salience and learning modulation are unified under the concept of **marker fields**, incorporating dopaminergic, noradrenergic, cholinergic, serotonergic and limbic influences [14], [15], [21], [22].

**4. Hyperengrams as global integrative states**
These provide the structural basis for conscious access, narrative continuity, working memory and self-consistency, integrating evidence from global workspace theory, default mode dynamics and network-level synchrony [16], [7], [17], [24], [25].

Together, these components define a **single algorithmic cycle**—Detect, Integrate, Mark, Execute—capable of explaining a wide range of cognitive and behavioural phenomena.

## 9.2 Theoretical Advantages of DIME

**1. Neural–computational unification**

DIME offers a representation of neural computation that is both **biologically grounded** and **computationally implementable**, resolving the long-standing divide between neuroscience descriptions and AI formalization.

**2. Multi-scale coherence**

Unlike models restricted to one level (spiking, neural populations, cortical modules, global networks), DIME spans **all relevant scales**, from millisecond-level dynamics to lifetime learning.

**3. Integration of emotion, value and cognition**

Marker systems allow emotion, motivation and subjective significance to be part of **computation itself**, not external additions.

**4. Unified account of memory, perception, planning and consciousness**

These traditionally separated domains appear naturally as **different instantiations** of the same DIME cycle.

**5. Algorithmic clarity**

DIME aims to articulate not only functional descriptions but also an explicit operational schema, enabling falsifiable predictions and implementable models.

## 9.3 Limitations of the Current Framework

Despite its integrative power, the present formulation of DIME has several limitations that require further empirical and computational refinement.

**1. Lack of a fully specified dynamical model**

The execution thread formalism requires a more detailed mathematical description (e.g., dynamical systems, Markov decision processes, neural ODEs).

**2. Marker system quantification**

While neurobiological grounding is strong, the exact mapping between neuromodulator levels, behavioural states and computational parameters requires further modeling [14], [15].

**3. Hyperengram mechanistic implementation**

Although consistent with global workspace and DMN theories, the operational mechanisms of hyperengram stabilization and dissolution are not yet fully formalized.

**4. Computational feasibility**

Large-scale simulation of engrams + threads + markers + hyperengrams remains computationally intensive with current infrastructure.

**5. Empirical validation**

DIME generates testable hypotheses, but large-scale neurophysiological validation is still pending.

These limitations point to rich opportunities for future investigations.

## 9.4 Testable Predictions

A major strength of DIME is its ability to generate **falsifiable predictions**, something often lacking in conceptual neuroscientific theories.

**Prediction 1: Multi-scale thread patterns**

Neural recordings should reveal **thread-like spatiotemporal motifs** that persist across modalities (vision, audition, memory, planning).

**Prediction 2: Marker-modulated thread selection**

Neuromodulator manipulations (dopamine, ACh, NE) should selectively alter **thread branching, persistence and merging**, not just local neural gain.

**Prediction 3: Hyperengram emergence under conscious perception**

Conscious access should correlate with the formation of **value-marked, large-scale integrative networks** rather than with local prediction-error suppression alone.

**Prediction 4: Engram dynamical repertoire**

Engrams should exhibit **context-dependent activation trajectories**, not fixed attractors.

**Prediction 5: Behavioural correlates**

Behaviour should reflect **thread-level computations**, such as multi-step internal simulation before overt action.

These predictions provide pathways for experimental and computational testing.

These predictions extend and operationalize the architectural consequences outlined in Section 4.5.

## 9.5 Implications for Future AI Systems

Section 8 outlined how DIME could transform artificial intelligence. Here we summarize the broader implications:

- AI systems will require **dynamic architectures**, not static computation graphs.
- Value must be **integrated intrinsically**, not externally imposed.
- World models must include **global integrators** (hyperengrams) to achieve narrative and goal coherence.
- Temporal reasoning should rely on **trajectory-level computation** rather than step-wise feedforward passes.
- Next-generation AI will likely be **marker-modulated**, **thread-based**, and **engram-structured**.

DIME thus provides a roadmap for the evolution of AI beyond deep learning and reinforcement learning.

## 9.6 Future Development of the DIME Framework

The next phases of the DIME program include:

**1. Mathematical formalization**

Developing full dynamical equations, plasticity rules and scaling laws.

**2. Computational simulation**

Implementing DIME-based artificial agents in controlled tasks (navigation, multi-step reasoning, embodied robotics).

**3. Neurobiological mapping**

Linking DIME components to empirically measurable brain signals (fMRI, ECoG, calcium imaging, laminar electrophysiology).

**4. Robotic implementation**

Embedding DIME in real-time control loops for humanoid robots or autonomous embodied agents.

**5. Integrative scientific synthesis**

Using DIME to unify research traditions across:

- predictive coding,
- attractor dynamics,
- reinforcement learning,
- global workspace theory,
- memory engram research,
- affective neuroscience.

### 9.7 Relation to the Companion Monograph

The present preprint provides a **conceptual and architectural summary** of the DIME framework. A complete elaboration—including mathematical derivations, diagrams, extended simulations, case studies and annexes—is provided in the companion monograph [1].

Researchers seeking deeper technical details, worked examples, or neurobiological mappings should refer to the monograph.

### 9.8 Concluding Remarks

DIME proposes that cognition emerges not from isolated modules but from the continuous interaction between dynamic representations (engrams), temporal processes (execution threads), intrinsic value systems (markers), and global integrative fields (hyperengrams). This perspective unifies perception, memory, prediction, learning, decision-making and consciousness under a single algorithmic cycle.

If validated empirically and implemented computationally, DIME may provide the foundation for a **new generation of cognitive architectures**, bridging the gap between biological intelligence and artificial systems.

### 9.6 Epistemic Positioning

DIME is not presented as a closed or definitive theory of brain function. Rather, it should be understood as a proposal for architectural compression — an attempt to reduce explanatory fragmentation by articulating a recurrent operational cycle that spans representation, dynamics, valuation and integration. Whether this compression reflects biological reality or merely provides a useful modeling abstraction remains an empirical question. Its value ultimately depends on whether it facilitates prediction, falsifiability and cross-domain integration more effectively than existing frameworks.

## APPENDIX 0 – Status of the arXiv Manuscript and Relation to the Companion Monograph

This arXiv manuscript is **not a partial or abbreviated version of the DIME theory**, but a **deliberately constrained architectural summary** whose sole purpose is to introduce, contextualize, and make accessible the core operational logic of the framework.

The **DIME theory itself is fully developed exclusively in the companion monograph**:

**The DIME Architecture - A New Unified Theory of the Brain**

I. C. Vladu, N. Bîzdoacă, I. Pirici, A. Bălșanu, E. N. Bondoc (2026), Zenodo. (Full DOI to be added)

The monograph constitutes the **single complete source of the theory**, including its formal definitions, mathematical structure, neurobiological grounding, algorithmic specification, simulations, extended comparisons, and applications.

**A. Purpose and Scope of the arXiv Manuscript**

The present arXiv version is intentionally limited in scope. It serves three functions only:

1. **Architectural exposition**
   To present the minimal operational commitments of the DIME framework at an abstract, system-level description.
2. **Conceptual orientation**
   To situate DIME relative to existing theories in neuroscience, cognitive science, artificial intelligence, and robotics.
3. **Explicit redirection to the monograph**
   To provide a clear map indicating where the full theory is developed.

This manuscript should therefore be read as a **conceptual and architectural gateway**, not as an independent or self-sufficient theoretical treatment.

**B. Structural Role of Tables 1–3**

To avoid ambiguity regarding scope and completeness, the architecture of the theory is explicitly stratified through Tables 1–3:

- **Table 1** lists the *minimal architectural commitments* assumed in this manuscript. These define the identity of the DIME framework at the operational level.
- **Table 2** summarizes the *main explanatory consequences* that logically follow from these commitments and are briefly outlined in the arXiv text.
- **Table 3** enumerates the *major conceptual, mathematical, biological, algorithmic, developmental, and philosophical components* that are **not developed in this manuscript** and are treated **exclusively and in full detail in the companion monograph**.

Together, these tables provide a **navigation structure**, explicitly distinguishing between:

- what is assumed,
- what is sketched,
- and what is intentionally omitted here.

No component listed in Table 3 is required to understand the arXiv manuscript, but **all are required to understand the DIME theory as a whole**.

**C. What Is Contained Exclusively in the Monograph**

The companion monograph provides the complete theoretical treatment, including but not limited to:

- full mathematical formalization of engrams, execution threads, marker fields, and hyperengrams;
- detailed neurobiological grounding across cellular, circuit, and network scales;
- formal algorithmic definitions and pseudocode for the DIME cycle;
- simulation results and computational experiments;
- extended comparisons with predictive coding, active inference, reinforcement learning, attractor dynamics, and global workspace theories;

- developmental, pathological, and philosophical analyses;
- AI and robotic implementations and design guidelines.

None of these elements are reproduced in full in the present manuscript.

**D. Epistemic Status**

The arXiv manuscript **does not claim theoretical completeness**, empirical exhaustiveness, or algorithmic closure. Its role is **expositional and integrative**, not demonstrative.

All claims regarding completeness, formal rigor, biological plausibility, and implementability of the DIME framework rest on the companion monograph.

Readers seeking the full theory are therefore explicitly directed to that work.

**E. Summary**

In short:

- **The monograph contains the full theoretical development.**
- **The arXiv manuscript is its architectural synopsis.**
- **Tables 1–3 define the boundary between the two.**

This separation is intentional and essential to the structure of the DIME research program.

**TABLE 1. Core architectural commitments of the DIME framework (present in arXiv)**

These commitments define the minimal operational architecture. All other concepts follow as consequences or extensions**.**

| No. | Core architectural commitment | Where in arXiv | Representative derived components (see Tables 2–3) |
|---|---|---|---|
| 1 | **Cognition as execution of spatiotemporal trajectories (not state transitions)** | Abstract; Sec. 1; 2.2 | Perception as pattern completion; memory re-execution; planning; imagination |
| 2 | **Engrams as distributed recurrent structures supporting multiple trajectories** | Abstract; 3.1 | Subengrams; reconsolidation; semantic abstraction; procedural stabilization |
| 3 | **Execution threads as the fundamental unit of neural process** | Abstract; 3.2; 4 | Decision as competition; internal simulation; narrative continuity |
| 4 | **Marker systems as intrinsic value-based control mechanisms** | Abstract; 1.1; 3.3 | Emotion, attention, motivation, learning modulation; regime switching |
| 5 | **Dual flow architecture: informational flow (FI) vs modulatory flow (FM)** | 1.2; 3.3 | Pre-rational selection; biasing without content transfer |
| 6 | **Hyperengrams as large-scale integrative states** | Abstract; 3.4 | Conscious episodes; working memory; global coherence |
| 7 | **Detect–Integrate–Mark–Execute as an invariant operational cycle** | Abstract; 4 | Unified mechanism for perception, memory, decision and action |

**TABLE 2. Major derived consequences outlined in arXiv (not axiomatic)**

These consequences follow logically from the core commitments but are not independent premises.

| No. | Derived domain | Key consequences | Where developed |
|---|---|---|---|
| 1 | Memory | No distinct memory “types”; memory as re-execution and reconsolidation | Sec. 3–5; Monograph Ch. 3, 5 |

| No. | Derived domain | Key consequences | Where developed |
|---|---|---|---|
| 2 | Decision-making | Decisions as competition between parallel execution threads | Sec. 3.2; 4; Monograph Ch. 4, 7 |
| 3 | Emotion | Emotion as control and valuation, not representational content | Sec. 3.3; Monograph Ch. 2, 5 |
| 4 | Attention | Attention as marker-driven stabilization of threads | Sec. 3.3; Monograph Ch. 2, 7 |
| 5 | Learning | Plasticity gated by marker state, not mere co-activation | Sec. 4.7; Monograph Ch. 5, 7 |
| 6 | Conscious access | Consciousness as stabilized hyperengram, not static workspace | Sec. 3.4; Monograph Ch. 5, 7 |
| 7 | Cognitive modes | Distinct cognitive regimes determined by marker configuration | Sec. 3.3; Monograph Ch. 7 |
| 8 | Development | Pre-rational marker-driven selection preceding deliberation | Sec. 1.1; Monograph Ch. 2 |
| 9 | Pathology | Mental disorders as architectural dysregulation | Mentioned; Monograph Ch. 5 |
| 10 | Large-scale networks | DMN, salience and control networks as emergent DIME configurations | Sec. 5; Monograph Ch. 5 |

```
          [ Hyperengram: global integration / consciousness ]
                  ↑                ↑                ↑
        -------------------------------------------------------------
        |                     Marker Systems                        |
        |    (dopamine, serotonin, ACh, NA, amygdala, PFC, etc.)    |
        |    regulate gain, plasticity, salience and selection      |
        -------------------------------------------------------------
                  ↑                ↑                ↑
     [ Engrams ] ←——— [ Execution Threads ] ———→ [ Actions ]
    (structural      (spatiotemporal neural programs      (motor,
     repertoire)        implementing cognition)       internal states)

DIME cycle:
DETECT → INTEGRATE → MARK → EXECUTE → (recurrent loop)

Core idea:
• Engrams provide structure.
• Threads provide process.
• Markers provide value and control.
• Hyperengrams provide global coherence.
```

**Figure 0.1. Conceptual schematic of the DIME architecture (textual abstraction)**

The present arXiv manuscript provides a conceptual and architectural overview of the DIME framework. Several structural, biological, algorithmic and developmental components of the theory are intentionally omitted for clarity and length. Appendix A explicitly lists the core concepts that are developed exclusively in the companion monograph, together with their location and theoretical relevance.

**TABLE 3. Conceptual commitments developed exclusively in the companion monograph**
These elements are not required to understand the arXiv paper but are essential for the full theory.

| No. | Conceptual commitment | Covers | Where in monograph | Why omitted from arXiv |
|---|---|---|---|---|
| 1 | Engram as executable neural program | Data–program duality; context-dependent execution | Ch. 1, 4 | Requires formal semantics |
| 2 | Subengrams and hierarchical composition | Feature units → concepts → narratives | Ch. 1 | Diagram-intensive |
| 3 | Execution thread formalism | State-space geometry; branching; competition | Ch. 4, 6 | Math-heavy |
| 4 | Marker field geometry | Vector space of value and significance | Ch. 2, 7 | Formal modeling |
| 5 | Marker-gated plasticity | Real-time learning during execution | Ch. 5, 7 | Biological detail |
| 6 | Hyperengram dynamics | Stabilization, dissolution, conscious continuity | Ch. 5, 7 | Requires temporal modeling |
| 7 | Developmental asymmetry (FM precedes cortex) | Ontogeny of valuation and control | Ch. 2 | Data-heavy |
| 8 | Pathology-specific mechanisms | PTSD, schizophrenia, depression | Ch. 5 | Clinical scope |
| 9 | Regime switching | Abrupt cognitive state transitions | Ch. 7 | Dynamical systems |
| 10 | Match / partial match / mismatch principle | Beyond prediction error | Ch. 4, 5 | Needs justification |
| 11 | Engram state-space topology | Attractors, basins, transitions | Ch. 6 | Visualization-heavy |
| 12 | Full DIME pseudocode | Scheduler, buffers, control logic | Ch. 6 | Too technical |
| 13 | Engram Machine (AI) | Cognitive architecture blueprint | Ch. 7, 10 | Applied focus |
| 14 | Robotic embodiment | Closed-loop physical instantiation | Ch. 10 | Hardware-specific |
| 15 | Philosophical implications | Agency, identity, free will | Ch. 12 | Outside arXiv scope |

## APPENDIX A (Informative) — Glossary of Core Concepts

This glossary provides concise definitions of terms central to the DIME architecture. It is intended to clarify terminology for interdisciplinary readers across neuroscience, AI, robotics and cognitive science.

**Engram**

A distributed, recurrent neural network encoding a **structured repertoire of possible activation trajectories**, supporting perception, memory, abstraction and planning.

**Engram Space ($\mathcal{E}$)**

The set of all engrame networks available to an agent, spanning multiple scales and modalities.

**Execution Thread**

A **spatiotemporal trajectory** of neural activation traversing engrams; the functional unit of cognitive processing.

**Thread Space ($\mathcal{T}$)**

The set of all currently active or latent execution threads.

**Marker System**

Distributed neuromodulatory and limbic mechanisms that **assign value, salience and contextual weighting** to engrams and threads.

**Marker Field (M(t))**

A vector-valued representation of instantaneous modulatory state influencing learning, excitability and selection.

**Hyperengram (H(t))**

A large-scale, temporally stable, value-marked network integrating **identity, context, goals and episodic history**, forming the basis of unified conscious experience.

**DIME Cycle**

The core algorithmic loop:

**Detect → Integrate → Mark → Execute**, operating continuously across scales.

# APPENDIX B — Mathematical Notes and Extended Formalization

The DIME framework can be extended mathematically to support simulation and theoretical analysis.

**B.1 Engram Structure**

Let each engram $E_i$ be defined as:

$$E_i = (V_i, A_i, T_i)$$

where:

- $V_i$= set of neurons/nodes;
- $A_i$= admissible activation patterns;
- $T_i: A_i \rightarrow A_i$= transition operator defining possible state trajectories.

Thus, engram dynamics are governed by:

$$x_{t+1} \in T_i(x_t)$$

Engrams are not fixed attractors, but **families of micro-trajectories**.

**B.2 Execution Threads as Directed Paths**

Let a thread $\tau$be:

$$\tau = \{(E_{i_1}, x_1), (E_{i_2}, x_2), \ldots, (E_{i_k}, x_k)\}$$

with transitions:

$$\left(E_{i_j}, x_j\right) \rightarrow \left(E_{i_{j+1}}, x_{j+1}\right)$$

Thread complexity is measured by:

- **length**,
- **branch factor**,
- **contextual weight**,
- **marker influence**.

**B.3 Marker Field Dynamics**

Marker field:

$$M(t) = \sum_{j=1}^{n} w_j \, S_j(t)$$

Plasticity or thread amplification can be modeled as:

$$g(t+1) = \sigma(\alpha g(t) + \beta M(t))$$

where $g(t)$ is gain.

**B.4 Hyperengram Formation**

Given active threads:

$$\mathcal{T}(t) = \{\tau_1, \tau_2, \ldots\},$$

hyperengram state defined as:

$$H(t) = \Phi(\mathcal{T}(t), M(t)),$$

where $\Phi$ is an integration function (pooling, synchronization, graph-based fusion, etc.).

This yields a dynamic state-space manifold for conscious access.

## APPENDIX C — Algorithmic Variants of the DIME Cycle

DIME is general enough to admit multiple implementations:

**C.1 Deterministic Variant**

Used for controlled environments:

*D = exact_match(S, E)*
*T = deterministic_update(T, D)*
*W = deterministic_marker(M, T)*
*O = deterministic_execute(T, W)*

**C.2 Probabilistic Variant**

Allows uncertainty:

$$p(\tau_{t+1}) = f(\tau_t, D(t), M(t))$$

**C.3 Reinforcement-Driven Variant**

Markers modulate expected return estimates:

$$Q(t+1) = Q(t) + \eta(M(t))(r - Q(t))$$

**C.4 Generative Simulation Variant**

Threads simulate futures:
*T_future = simulate(T, E, M)*

# APPENDIX D — Experimental Predictions and Suggested Neuroscientific Tests

**Prediction D1 — Thread Continuity Across Modalities**
Evidence: long-range sequential replay in hippocampus & cortex.
**Test:** Multi-area electrophysiology showing cross-modal sequential propagation.

**Prediction D2 — Marker-Gated Thread Selection**
Evidence: dopamine modulates exploratory behaviour.
**Test:** Pharmacological manipulation altering branching structure of neural trajectories.

**Prediction D3 — Hyperengram Formation Linked to Conscious Access**
Evidence: global workspace activation patterns.
**Test:** fMRI/ECoG showing global integration only when reportable perception occurs.

**Prediction D4 — Context-Dependent Engram Dynamics**
Evidence: reconsolidation.
**Test:** Same neural ensemble shows different trajectories depending on task or internal state.

# APPENDIX E — Computational Implementation Guidelines

To implement DIME in simulation or AI:

**E.1 Representing Engrams**

- Use recurrent networks with dynamic attractor basins.
- Encode multiple potential activation trajectories.

**E.2 Implementing Threads**

- Use path-following mechanisms (RNNs, transformers with recurrence, neural ODEs).
- Allow branching and merging via gating.

**E.3 Marker Systems**

- Represent neuromodulation as global vectors affecting gain, learning rate, exploration.
- Implement affective tagging and priority mechanisms.

**E.4 Hyperengram Construction**

- Use global latent vectors or memory pools integrating multi-thread summaries.
- Update each cycle to maintain narrative coherence.

**E.5 Real-Time DIME Simulation**

- Implement the DIME loop asynchronously to mimic biological temporal structure.

## APPENDIX F — Suggested Robotic Architectures Based on DIME

A DIME-inspired embodied agent would include:

**1. Sensorimotor Engrams**

- Learned recurrent representations of sensor states.

**2. Execution Threads for Planning**

- Hierarchical motion planning as thread continuation.

**3. Marker Field Coupled to Body States**

- Battery → dopamine analog;
- Motor load → noradrenaline analog;
- Thermal load → serotonin analog.

**4. Hyperengram for Context and Goals**

- High-level state integrating task context, past trajectories, and planning horizon.

**5. Closed-Loop DIME Control**

*Detect → sensory encoding*
*Integrate → thread update*
*Mark → modulated priority*
*Execute → action + forward simulation*
This architecture supports self-stabilizing, adaptive behaviour.

## APPENDIX G — Relationship Between DIME and the Companion Monograph

The companion monograph [1] expands all components of the DIME framework:

**Included in the monograph:**

- full mathematical derivations of thread dynamics,
- extended neurobiological evidence,
- diagrams and illustrations for the full architecture,
- detailed pseudocode and algorithmic variants,
- simulation results for DIME-inspired agents,
- annexes covering theory integration (predictive coding, RL, GWT, etc.).

**Role of this Appendix**

The present appendix focuses on conceptual and computational supplements necessary to support the preprint, with references to areas where the monograph provides deeper elaboration.

## Author Contributions

**I.C. Vladu** conceived and developed the DIME framework in its entirety, including the conceptual architecture, theoretical formulation, algorithmic design, mathematical structuring, and the full writing of the manuscript. He coordinated the integration of interdisciplinary perspectives and ensured coherence between the conceptual, biological, mathematical, and computational components.

**N. Bîzdoacă** contributed domain-specific expertise in robotics and artificial intelligence, providing technical validation regarding the implementability of the DIME architecture in artificial agents and robotic control systems. He offered interdisciplinary feedback connecting the theoretical framework to engineering constraints.

**I. Pirici** and **A. Bălșeanu** contributed domain-specific expertise in neuroanatomy and clinical neuroscience. They evaluated the biological plausibility of the proposed mechanisms, ensured that the neuroanatomical interpretations align with current evidence, and validated cross-scale correspondences between empirical neuroscience and the DIME abstractions.

**E. N. Bondoc** contributed clinical psychology and neuropsychiatry expertise, providing evaluation of the cognitive and behavioural plausibility of DIME. He reviewed the alignment between the framework and observed clinical phenomena (memory, attention, affective modulation, and coherence of conscious states), ensuring compatibility with psychological and psychiatric evidence.

**All authors** discussed the framework, provided domain-level validation within their areas of expertise, and approved the final version of the manuscript.

## Funding

This research received no external funding.

## Competing Interests

The authors declare no competing interests.

## Acknowledgments

The authors thank the interdisciplinary colleagues from UCV and UMF Craiova for discussions that helped refine several aspects of the DIME architecture.